# Thermoelectric Band Engineering:
# The Role of Carrier Scattering

Evan Witkoske, Xufeng Wang, and Mark Lundstrom
Purdue University, West Lafayette, IN USA

Vahid Askarpour and Jesse Maassen
Dalhousie University, Halifax, Nova Scotia, CA

**Abstract –** Complex electronic band structures, with multiple valleys or bands at the same or similar energies can be beneficial for thermoelectric performance, but the advantages can be offset by inter-valley and inter-band scattering. In this paper, we demonstrate how first-principles band structures coupled with recently developed techniques for rigorous simulation of electron-phonon scattering provide the capabilities to realistically assess the benefits and trade-offs associated with these materials. We illustrate the approach using n-type silicon as a model material and show that intervalley scattering is strong. This example shows that the convergence of valleys and bands can improve thermoelectric performance, but the magnitude of the improvement depends sensitively on the relative strengths of intra- and inter-valley electron scattering. Because anisotropy of the band structure also plays an important role, a measure of the benefit of band anisotropy in the presence of strong intervalley scattering is presented.

## 1. Introduction

The performance of a thermoelectric device is controlled by the material figure of merit [1-4],

$$zT = \frac{S^2 \sigma T}{\kappa_e + \kappa_L}, \tag{1}$$

where $S$ is the Seebeck coefficient, $\sigma$ the electrical conductivity, $T$ the absolute temperature, $\kappa_e$ the electronic thermal conductivity, and $\kappa_L$ the lattice thermal conductivity. How various material parameters affect $zT$ is well understood (e.g. [1-4]), but parabolic energy band analyses suggest that the prospects for improving the electronic contribution to $zT$ are limited [5]. Indeed, much of the recent progress in increasing $zT$ has been achieved by lowering the lattice thermal conductivity (e.g. [6-11]). There is, however, currently considerable interest in examining complex thermoelectric materials, which may provide improved electrical performance not possible with simple parabolic energy bands (e.g. [5,9,12-17]). First-principles calculations of thermoelectric transport parameters are routinely performed to assess complex thermoelectric materials [18, 19], but the treatment of electron scattering greatly complicates the analysis leading to the widespread use of rigorous band structures coupled with a highly simplified treatment of scattering – the constant relaxation time approximation (CRTA). The recent development of techniques to rigorously compute scattering rates [20-24] presents an opportunity to include detailed band structure and scattering physics in the analysis of complex thermoelectric materials. Our goal is to illustrate the importance of going beyond the CRTA by presenting calculations for n-type silicon, which has an anisotropic, multi-valley conduction band, as a model material.

Equation (1) can be re-expressed as

$$zT = \frac{S'^2}{L' + 1/b} ,\qquad(2)$$

where $S' = S/(k_B/q)$ is the dimensionless Seebeck coefficient, $L' = L/(k_B/q)^2$ the dimensionless Lorenz number, and

$$b \equiv \frac{\sigma T}{\kappa_L (k_B/q)^2} .\qquad(3)$$

Assuming unipolar conduction, the conductivity can be written as $\sigma = nq\mu_n$, where $n$ is the carrier density and $\mu_n$ the mobility. If we further assume parabolic energy bands, then we can write

$$b = B\mathcal{F}_{1/2}(\eta_F) \qquad(4)$$

where

$$B \equiv \frac{N_V}{4}\left(\frac{2m_V^* k_B T}{\pi \hbar^2}\right)^{3/2} \frac{q\mu_n T}{\kappa_L}\left(\frac{k_B}{q}\right)^2 .\qquad(5)$$

In (5)

$$\eta_F = (E_F - E_C)/k_B T ,\qquad(6)$$

is the dimensionless Fermi energy (chemical potential), $\mathcal{F}_{1/2}(\eta_F)$ is the Fermi-Dirac integral of order $j=1/2$ as written in the Blakemore form [25]

$$\mathcal{F}_j(\eta_F) = \frac{1}{\Gamma(j+1)} \int_0^\infty \frac{\eta^j d\eta}{1+e^{\eta-\eta_F}} , \qquad (7)$$

and $m_v^*$ is the DOS effective mass of a single valley, and $N_V$ is the valley degeneracy. Note that the b-factor in (3) can be evaluated for any band structure while the B-factor in (5) assumes parabolic energy bands.

The quantity, $B$, is the "material factor" $\beta$ introduced by Chasmar and Stratton [26]. It was discussed extensively by Mahan, who called it the "B-factor" [4]. The important role it plays in thermoelectric material design has been recently discussed by Wang et al. [13], who call $B$ the quality factor. The B-factor is, however, not the whole story. For example, recent work has focused on identifying complex thermoelectric materials with increased Seebeck coefficient (e.g. [12]) or reduced Lorenz number (e.g. [27]). While there are many trade-offs involved in thermoelectric material design, our focus in this paper is on how multiple valleys affect the b-factor. We do so using rigorous treatments of band structure and electron-phonon scattering.

Equation (5) suggests that materials with many degenerate valleys will be good thermoelectrics. Mahan points out that good thermoelectric materials are all multi-valley semiconductors [4] (but n-GaAs, a single valley material, also shows promise [28].) Recent work on thermoelectric band engineering has focused on engineering materials to achieve a large number of valleys and/or bands near the Fermi level (e.g. [14, 15]). As written, however, (5) does not highlight the trade-off involved in increasing the valley/band degeneracy. More valleys and bands provide more states to which carriers can scatter. Increasing $N_V$ should decrease the scattering time and lower the mobility. These considerations have been discussed by Wang et al. [13], who argue that intra-valley scattering typically tends to dominate, so increasing $N_V$ should increase $B$. A recent study using analytical descriptions of energy bands and scattering processes concluded that multiple valleys may or may not be beneficial depending on the material-dependent specifics of inter-valley scattering [29]. First-principles calculations of thermoelectric transport parameters allow complex band structures to be treated without approximation, but since they commonly make the constant relaxation time approximation, such simulations cannot answer how much increasing the number of valleys improves the b-factor. Rigorous calculations of electron-phonon scattering rates are, however, now possible. In this paper, we show that the capabilities now exist to more thoroughly address the question of how multiple valleys affect thermoelectric performance.

The paper is organized as follows. The equations for the thermoelectric transport coefficients are summarized in Sec. 2; the goal of the paper is to solve these equations and assess the impact of intervalley scattering on the b-factor as given by (3). In Sec. 3, we solve the thermoelectric equations for parabolic band semiconductors and use the solutions to illustrate issues that are examined with a full numerical band structure and first principles scattering rates in Sec. 4. To illustrate the rigorous treatment of an anisotropic, multi-valley semiconductor, we consider n-type Si, which has six conduction band valleys. We use a DFT-generated band structure along with electron-phonon scattering rates informed by rigorous simulations to compare six-valley silicon to corresponding single spherical band structures. The results will show that for Si the benefits of the six multiple valleys are largely offset by intervalley scattering, but the anisotropic band structure does provide benefits over a simple, isotropic band structure. To understand the results presented in Sec. 4, we must separate the effects of intra- and inter-valley scattering from those due to the anisotropy of the band structure. In Sec. 5, we examine two simple metrics that can be used to assess the thermoelectric potential of complex band structures. The paper concludes with a Summary in Sec. 6.

## 2. Approach

The expressions for the thermoelectric transport coefficients that result from a relaxation time approximation solution to the Boltzmann Transport Equation (BTE) are:

$$\sigma = \int_{-\infty}^{+\infty} \sigma'(E) dE \tag{8a}$$

$$S = -\frac{1}{qT} \frac{\int_{-\infty}^{+\infty}(E-E_F)\sigma'(E)dE}{\int_{-\infty}^{+\infty}\sigma'(E)dE} \tag{8b}$$

$$\kappa_0 = \frac{1}{q^2 T} \int_{-\infty}^{+\infty} (E-E_F)^2 \sigma'(E) dE \tag{8c}$$

$$\kappa_e = \kappa_0 - T\sigma S^2, \tag{8d}$$

where the differential conductivity is

$$\sigma'(E) = q^2 \Xi(E)\left(-\frac{\partial f_0}{\partial E}\right), \tag{8e}$$

and the transport distribution is [30]

$$\Xi_{ij}(E) \equiv \frac{1}{\Omega}\sum_{\vec{k}}\left(v_i \tau_{jk} v_j\right)\delta(E-E_k), \tag{8f}$$

where $\Omega$ is a normalization volume. Next, we assume a diagonal transport distribution tensor and write the transport distribution in Landauer form as [31]

$$\Xi(E) = \frac{2}{h}\left(M(E)/A\right)\lambda(E), \tag{8g}$$

where $M(E)/A$ is the number of channels for conduction per unit cross-sectional area vs. energy. (See Appendix A for a derivation of (8g).) We compute $M(E)/A$ from a DFT-generated band structure using the open source tool, LanTrap 2.0 [32]. The energy-dependent mean-free-path for backscattering is also needed; it is defined as [31]

$$\lambda(E) \equiv 2\frac{\langle v_x^2 \rangle}{\langle |v_x| \rangle}\tau_m(E), \tag{9}$$

where $\langle v_x^2 \rangle / \langle |v_x| \rangle$ is an energy dependent angle-averaged velocity and is computed from the DFT-generated band structure. For acoustic deformation potential (ADP) scattering in the elastic limit, the scattering rate is isotropic, equal to the momentum relaxation rate, and proportional to the DOS:

$$\frac{1}{\tau(E)} = \frac{1}{\tau_m(E)} \propto K_{el-ph} DOS(E), \tag{10}$$

which can be computed directly from the numerical band structure. The electron-phonon coupling parameter, $K_{el-ph}$, is proportional to the deformation potential squared. The electron-phonon scattering rates in non-polar semiconductors generally follow the DOS [33]. The rigorously computed electron-phonon scattering rates presented in Sec. 4 confirm that for silicon, the scattering rate follows the DOS, so for the numerical calculations presented in Sec. 4, we take $K_{el-ph}$ from the rigorously computed scattering rate. We will refer to scattering described by (10) as "DOS scattering." As discussed next, for parabolic energy bands and simple scattering processes, equations (8) simplify and can be solved analytically.

## 3. Results: Parabolic bands

To illustrate how multiple, anisotropic valleys with and without intervalley scattering affects the b-factor, we present some calculations for parabolic energy bands. For parabolic bands [31, 34],

$$M(E)/A = \frac{m_\sigma^*}{2\pi\hbar^2}(E - E_C), \qquad (E - E_C) > 0 \tag{11}$$

where $m_\sigma^*$ is the "distribution of modes" effective mass [28] (for more details on the "distribution of modes effective mass," please see the Supplementary material). It is analogous, but not equal to, the conductivity effective mass in the traditional formulation. For spherical energy bands, $m_\sigma^* = N_V m^*$, where $N_V$ is the number of valleys and $m^*$ is the effective mass of each valley. For the ellipsoidal conduction band of Si [31]

$$m_\sigma^* = 2m_t + 4\sqrt{m_t m_l}. \tag{12}$$

For spherical energy bands, the MFP for backscattering is [34]

$$\lambda(E) = \frac{4}{3} v(E) \tau_m(E), \tag{13}$$

which can be written in power law form as

$$\lambda(E) = \lambda_0 \left( (E - E_C)/k_B T \right)^r, \qquad (E - E_C) > 0 \tag{14}$$

where $r$ is a characteristic exponent. For DOS scattering $r = 0$, and the MFP is independent of energy. For a constant scattering time (CRTA), $r = 1/2$. By using (11) and (14) in (8), we find [34]

$$\sigma = \frac{2q^2}{h} \lambda_0 \left( \frac{m_\sigma^* k_B T}{2\pi \hbar^2} \right) \Gamma(r+2) \mathcal{F}_r(\eta_F) \tag{15a}$$

$$S = -\left( \frac{k_B}{q} \right) \left( \frac{(r+2) \mathcal{F}_{r+1}(\eta_F)}{\mathcal{F}_r(\eta_F)} - \eta_F \right) \tag{15b}$$

$$\kappa_0 = T \left( \frac{k_B}{q} \right)^2 \frac{2q^2}{h} \lambda_0 \left( \frac{m_\sigma^* k_B T}{2\pi \hbar^2} \right) \times \\ \left[ \Gamma(r+4) \mathcal{F}_{r+2}(\eta_F) - 2\eta_F \Gamma(r+3) \mathcal{F}_{r+1}(\eta_F) + \eta_F^2 \Gamma(r+2) \mathcal{F}_r(\eta_F) \right], \tag{15c}$$

from which the b-factor can be computed if we assume a lattice thermal conductivity, $\kappa_L$, and scattering parameters, $\lambda_0$ and $r$. We assume a lattice thermal conductivity of $\kappa_L = 1$ W/m-K. All calculations are done at 300 K.

### 3.1 Results: Spherical, parabolic bands and multiple valleys

We begin with spherical bands and consider two cases; the first assumes only intra-valley scattering and the second assumes equally strong intra- and inter-valley scattering. We assume $m_\sigma^* = N_V m_0$ and vary the valley degeneracy from $1 \leq N_V \leq 10$.

When only intra-valley scattering is assumed, a MFP of $\lambda = \lambda_0 = 10\,\text{nm}$, which is independent of $N_V$. The second case assumes equally strong intra- and inter-valley scattering, so the MFP goes as $\lambda = \lambda_0/N_V$. For every value of $N_V$ the maximum $zT$ is found by sweeping the Fermi level to find $\hat{\eta}_F$ at the maximum $zT$. Equation (11a) then gives $\sigma(\hat{\eta}_F)$ and from (3), $b(\hat{\eta}_F)$ is computed. Using (4), we can then deduce $B = b(\hat{\eta}_F)/\mathcal{F}_{1/2}(\hat{\eta}_F)$.

The results are shown in Fig. 1. To understand these results note that according to (15a) $\sigma \propto m_\sigma^* \lambda$. When there is only intra-valley scattering, $\sigma \propto (N_V m_0)\lambda_0$, so the conductivity, and therefore the B-factor increases linearly with the number of valleys. The increase of the b-factor with $N_V$ is however sub-linear. This behavior is due to the fact that the Fermi level that maximizes $zT$ moves down as $N_V$ increases, causing the conductivity to increase sub-linearly with $N_V$. This results in the b-factor rolling over despite the fact that the Seebeck coefficient increases with $N_V$. Although neglecting inter-valley scattering may seem artificial, it does have some relevance to nanostructured thermoelectric materials in the "small nanostructure size" (SNS) limit [10,11]. In this limit, the MFP is determined by grain size and independent of $N_V$ or $m^*$. Under these conditions, a large $N_V$ (or equivalently a large effective mass in a single valley) is beneficial.

When the intra and inter-valley scattering rates are equal, Fig. 1 shows that the B- and b-factors are independent of $N_V$. This follows directly from $\sigma \propto m_\sigma^* \lambda = (N_V m_0)(\lambda_0/N_V)$, being independent of the number of valleys when inter-valley scattering is strong. Figure 1 illustrates the importance of accurately quantifying the relative strengths of intra- and inter-valley scattering. Figure 1 also shows the difference between the b- and B-factors. Since the B-factor is well-defined only for parabolic energy bands while the b-factor can be computed for arbitrary band structures, the b-factor may be more relevant for treating complex thermoelectric materials. It is also more directly related to measured quantities.

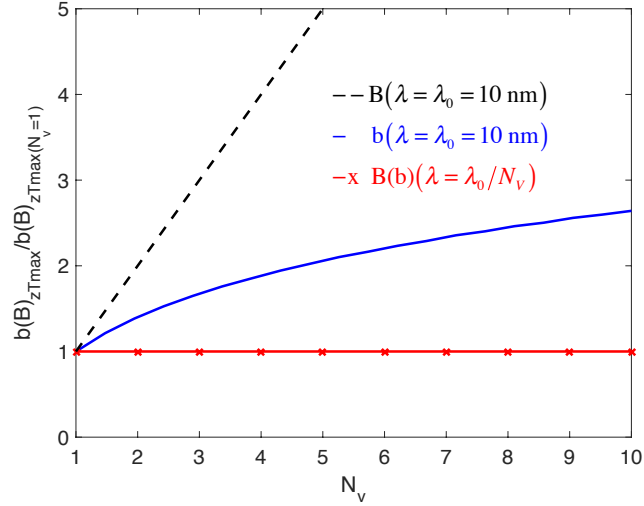

Fig. 1  The b- and B-factors versus the number of degenerate valleys. The solid line is the b-factor assuming intra-valley scattering only (i.e. a constant MFP). The dashed line is the B-factor assuming a constant MFP. The solid line with x's is the result for both the b- and B-factors assuming equally strong intra- and inter-valley scattering. In all cases, the results are normalized to one when the number of valleys is one.

### 3.2 Results: Silicon-like Anisotropic Valleys

Complex thermoelectric materials often have anisotropic band structures that can boost thermoelectric performance [17]. Both scattering and anisotropy affect the performance of a thermoelectric material. In this section, we use a silicon-like conduction band with six equivalent ellipsoidal bands to examine how anisotropy affects the b-factor. Three cases are considered: i) a constant MFP, ii) a constant scattering time, iii) a scattering time inversely proportional to the total DOS of all valleys. The first two cases are considered because they are commonly used assumptions. Case iii) corresponds to equally strong intra- and inter-valley scattering. The results illustrate the connection between band anisotropy and scattering.

To treat this problem, we must extend (13) and (14). Details can be found in the supplementary material; the results are:

Case i) Constant MFP independent of energy and effective mass, $r = 0$:

$$\lambda = \lambda_0 = 10 nm \qquad (16a)$$

Case ii) Constant scattering time, $r = 1/2$,

$$\lambda(E - E_c) \propto \sqrt{(E - E_c)} \left( \frac{1/m_\ell + 2/m_t}{1/\sqrt{m_\ell} + 2/\sqrt{m_t}} \right) \tau_0 \qquad (16b)$$

Case iii) Equally strong intra- and inter-valley DOS scattering,

$$\lambda \propto \frac{\lambda_0}{6} \left( \frac{m_t + 2m_\ell}{m_t^2 m_\ell + 2(m_t m_\ell)^{3/2}} \right). \qquad (16c)$$

In case ii), the constant scattering time $\tau_0$ is set using $\mu = q\tau_0 / m_c^*$ with the mobility and conductivity effective mass from silicon of 1360 cm$^2$/Vs and 0.26 $m_0$ [35 (pg.166)] respectively. In case iii), the MFP is scaled so that, $\lambda(m_\ell = 0.93 m_0, m_t = 0.19 m_0) = 10 nm$. In each case, the transverse effective mass $m_t/m_0$ is varied with $m_\ell/m_0$ held constant at $m_\ell/m_0 = 0.93$. For every $m_t/m_0$ value, the Fermi level is swept to find $\hat{\eta}_F$ at the maximum $zT$. Equation (15a) then gives $\sigma(\hat{\eta}_F)$ and from (3), $b(\hat{\eta}_F)$ is computed. Using (5), we then compute $B = b(\hat{\eta}_F) / \mathcal{F}_{1/2}(\hat{\eta}_F)$. Only the b-factor will be shown, because the B-factor displays the same trends.

The results are shown in Fig. 2. For the constant MFP case, the b-factor continues to increase with increasing $m_t$ (decreasing anisotropy). This behavior is similar to the constant MFP case shown in Fig. 1 and occurs because $\sigma \propto m_\sigma^*$ when the MFP is constant, and according to (12), $m_\sigma^*$ increases as $m_t$ increases. For a constant scattering time, Fig. 2 shows that the b-factor increases more slowly with increasing $m_t$. This occurs because an increasing $m_t$ increases $m_\sigma^*$, but it decreases the velocity so the MFP decreases with increasing $m_t$ according to (16b). In case iii), however, the trend is opposite to that of cases i) and ii). Even though $m_\sigma^*$ increases as $m_t$ increases, the MFP decreases rapidly and continues to do so with increasing $m_t$ according to (16c). This occurs because the increasing $m_t$ lowers the velocity and increases the DOS, which lowers the scattering time. Because the MFP is the product of velocity and scattering time, it decreases rapidly within increasing $m_t/m_0$. In the end, cases i) and ii) benefit from a larger transverse effective mass, while case iii) is maximum for a transverse effective mass approaching zero. The large differences in Fig. 2 at $m_t/m_0 = 0.19$ (the Si value) illustrate the importance of a proper treatment of scattering.

The fact that different types of scattering and anisotropies affect thermoelectric performance is well understood (e.g. [3], [8,9,12]), but the interplay of scattering, anisotropy, and valley degeneracy is not well understood [29]. In the next section, we examine scattering in anisotropic, multi-valley semiconductors by using a DFT-generated silicon band structure and rigorously computed electron-phonon scattering rates.

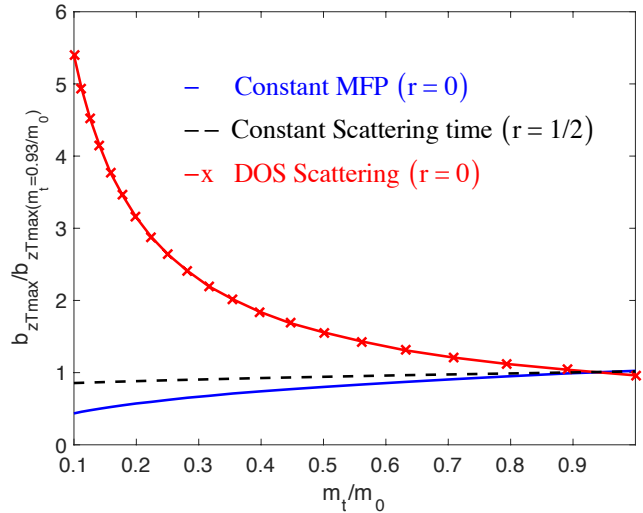

Fig. 2 Normalized b-factor vs. increasing $m_t^*/m_0$ (decreasing anisotropy). The solid line assumes a constant MFP, case i). The dashed line assumes a constant scattering time, case ii), and the solid line with x's, assumes a scattering rate proportional to the total DOS, case iii). In all cases the b-factors are normalized to one when the valleys are spherical with $m_t = m_\ell$.

## 4. Results: Numerical bands and scattering rates

Thermoelectric materials generally have complex band structures that can only be described numerically. As illustrated here for the relatively simple case of silicon, an assessment of the performance potential of a material cannot be done without a careful consideration of how band structure affects electron scattering. The main challenge in doing such calculations is the specification of $\lambda(E)$. As discussed below, rigorous, first-principles calculations of the electron scattering time [20-24, 36] can be used to determine $\lambda(E)$.

### 4.1 Numerical calculation of scattering rates

The numerical calculations were done as follows. The structural relaxation, self-consistent ground state, and DFPT calculations were performed with the Quantum Espresso package [37], using Perdew-Zunger LDA exchange-correlation, norm-conserving pseudo-potential, a 48 Ry plane wave energy cutoff, and a 16×16×16 Monkhorst-Pack *k*-mesh. The converged lattice constant is 5.38 Å in agreement with similar DFT studies [24]. The scattering rate calculations were performed with the EPW package [20, 21] to extract both the relaxation time and the momentum relaxation time at 300 K [22]. Electronic energies, phonon energies, and the electron-phonon matrix elements were initially calculated for zone-centered 6×6×6 coarse *k*- and *q*-grids. Eight maximally localized Wannier functions [36] of sp$^3$ symmetry were generated and serve as the basis to transform the electronic Hamiltonian, phonon dynamical matrix, and el-ph coupling Hamiltonian. From the Wannier representation, the electron and phonon energies and el-ph coupling matrix elements are interpolated onto dense *k*- and *q*-grids of 60×60×60 and used to calculate the scattering rate. A Gaussian smearing parameter of 30 meV was used for energy conservation. To reduce computation time, only those *k*-points in the irreducible wedge of the Brillouin zone were included in the analysis. The carrier concentration, determined from the Wannier representation density-of-states (DOS) [36], was chosen as 10$^{15}$ cm$^{-3}$. To distinguish intra-valley and inter-valley scattering processes, inter-valley transitions were identified when the change in the electron wave vector was $|q| > 0.505$ Å$^{-1}$ (0.25 in reduced coordinates) or when there was a change in band number when going from initial to final band. All other transitions were categorized as intra-valley scattering.

Figure 3 shows the computed scattering rate vs. energy for electrons in the conduction band of silicon. Each value plotted represents a distinct point in k-space. Figure 3 shows that both the scattering rate (points) and momentum relaxation rate (+) follow the DOS (line). The fact that the momentum relaxation rate and scattering rate are basically equal indicates that electron-phonon scattering is isotropic in silicon, as expected [35]. These numerical calculations confirm that for n-Si, $1/\tau_m(E) \propto DOS(E)$, as expected from simpler, phenomenological treatments [33, 35].

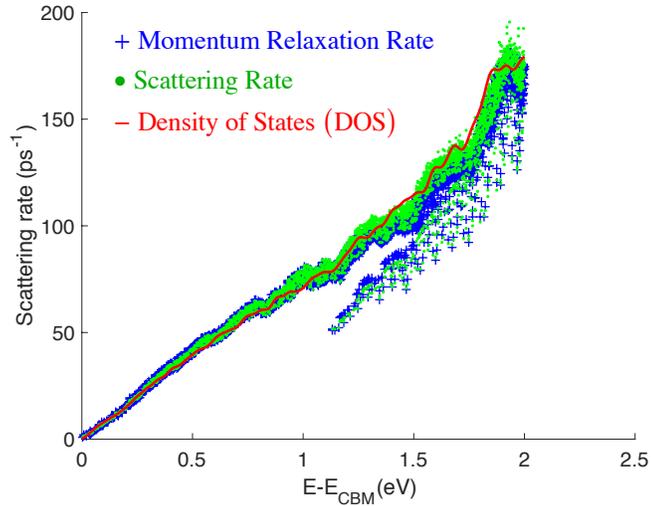

Fig. 3  Scattering rate vs. energy for electrons in the conduction band of silicon. Each point represents a point in k-space. The results show that the scattering rate (points) depends primarily on the energy of the state and not its location in k-space. The + symbols are the momentum relaxation rate, which is essentially the same as the scattering rate in silicon. Both the scattering rate and momentum relaxation rate follow the density-of-states (line), especially around the energy most relevant for transport, $E - E_C \ll 1$ eV.

The next question concerns the relative importance of intra- vs. inter-valley scattering. As discussed earlier, the benefits of multiple valleys are reduced if inter-valley scattering dominates. A phenomenological treatment indicated that both intra- and inter-valley scattering are important in n-Si [38]. Figure 4, which compares numerically calculated intra-valley and inter-valley scattering rates for electrons in the conduction band of silicon, shows that inter-valley scattering dominates in this material. Near the band edge, which is what matters for the thermoelectric coefficients, the inter-valley scattering rate per valley is comparable to that of intra-valley, i.e. there is equal probability to scatter to any valley. The numerical calculations also provide the room temperature, phonon-limited mobility in bulk Si as $\mu_n = 1480$ cm$^2$/V-s, which is within 10% of the experimental value of 1360 cm$^2$/V-s [38] and suggests that the calculations are reliable. Finally, the calculations also provide a rigorous solution to the BTE, which shows that the RTA, which is assumed in eqns. (8), is accurate to within a few percent.

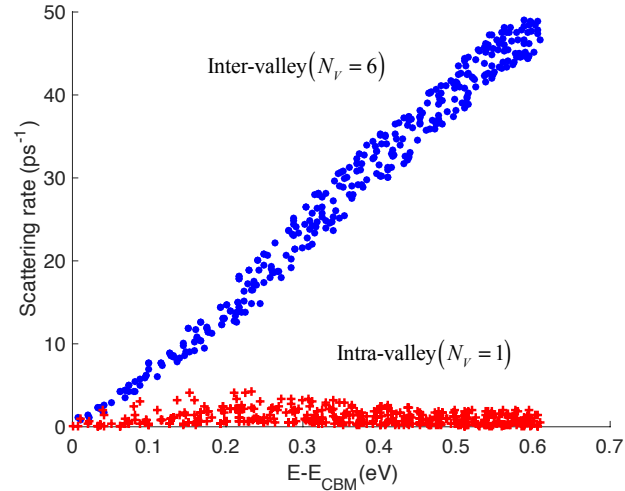

Fig. 4  Scattering rate vs. energy for electrons in the conduction band of silicon. In the plot, the + symbols show the intra-valley scattering rate and the points show the inter-valley scattering rate. Near the band edge, i.e. $E < 0.1$ eV there is roughly the same probability to scatter within a valley as to scatter to a different valley.

**4.2 Calculation of the b-factor**

When discussing the benefits of multiple, anisotropic valleys, the point of comparison is a corresponding (in some appropriate sense) spherical valley semiconductor. The calculations presented next will compare the performance of the n-Si using a rigorous band structure with corresponding spherical models. From the rigorous calculations shown in Fig. 3, we extract the electron-phonon coupling parameter, $K_{el-ph}$ needed in (10). Equations (8) can then be solved using (9) and (10) to specify the energy-dependent MFP. The distribution of channels, $M(E)$, is extracted directly from the band structure using LanTrap 2.0 [32], and the angle-averaged velocity, $\langle v_x^2 \rangle / \langle |v_x| \rangle$, which is needed for $\lambda(E)$, is also computed directly from the band structure. With this information, (8) can be solved repeatedly as the Fermi level is swept to find $\hat{\eta}_F$ at the maximum $zT$. As in the parabolic band case, for these model calculations, we assume $\kappa_L = 1$ W/m-K rather the actual value for Si. This procedure provides $\sigma(\hat{\eta}_F)$, the conductivity at peak $zT$, and from (3), $b(\hat{\eta}_F)$ is computed. Since the b-factor is well-defined for general band structures, but the B-factor only applies to parabolic band semiconductors, we will focus on the b-factor.

Table 1 shows the computed b-factors for several different cases. Case A1 uses the full, numerical Si conduction band and assumes equally strong intra- and inter-valley scattering. Case A2 assumes only intra-valley scattering with $K_{el-ph}$ being replaced by $K_{el-ph}/6$. This example enjoys the benefits of six valleys without suffering from increased scattering between valleys, therefore the b-factor should be larger. Table 1 shows that the b-factor is significantly larger when inter-valley scattering is ignored (case A1 vs. A2).

When complex band structures are anisotropic in the right way, they can boost $zT$ even when scattering between valleys occurs. Case B1 in Table 1 displays the b-factor for a corresponding spherical band. The effective mass of this spherical band is chosen to give the same $M(E)$ as the full, numerical band. Because the Si conduction band is nearly parabolic, the numerically extracted $M(E)$ closely follows the analytical expression for parabolic bands, (11). From the numerical $M(E)$, we extract $m_\sigma^*$ and then set the effective mass of the corresponding spherical band to $m^* = m_\sigma^*$. This procedure would give the same conductivity in Cases A1 and B1, if the MFP's in the two cases were the same.

The procedure described above produces, however, a different MFP, $\lambda(E)$, from case A1 for two reasons. The first reason is that this process produces different DOS's for the two cases, and, therefore, different scattering times according to (10). The second reason is that the velocity ratio, $\langle v_x^2 \rangle / \langle |v_x| \rangle$ in the MFP expression, (9), is smaller for the corresponding spherical band because the benefits of anisotropy are absent. Since we are interested in ascertaining the benefits of anisotropy in Case B1, we assume the same scattering times as in Case A1, but instead of using the numerically calculated $\langle v_x^2 \rangle / \langle |v_x| \rangle$, we use the smaller value for a spherical band, $\langle v_x^2 \rangle / \langle |v_x| \rangle = 2\sqrt{2(E-E_C)/m^*}/3$. Comparing the b-factor in Case B1 of Table 1 to Case A1, we see that anisotropy increases the b-factor by a factor of 1.24 for the case of n-Si.

Finally, case B2 assumes a spherical, parabolic energy band with an effective mass chosen to give one-sixth of the $M(E)$ as the full, numerical band. While case B1 represents all six conduction band valleys with a single spherical energy band, case B2 represents only a single conduction band valley with a spherical energy band. In this case, we assume a scattering time that is six times longer than for case A1, to account for the fact that there is no inter-valley scattering in case B2. In comparison to case B2, case A1 benefits from six times as many valleys and from valley anisotropy, but it suffers from six times as much scattering. Comparing the b-factor in case B2 of Table 1 to case A1, we see that the case A1 b-factor is 2.6 times that of case B2. The benefit of the six valleys is offset by six times more scattering, and the improvement is due to valley anisotropy.

In summary, for n-Si, the advantages of the six conduction band valleys are offset by the increased scattering of electrons between the valleys. The anisotropy of the valleys, however, provides a light effective mass in the direction of transport, which increases the b-factor in comparison to a corresponding spherical valley. To assess the potential of a thermoelectric material, the beneficial effects of valley anisotropy and valley degeneracy must be weighed against the detrimental effects of inter-valley/band scattering.

| Case | Band structure | Scattering | $b(\hat{\eta}_F)$ | $\hat{\eta}_F$ |
|---|---|---|---|---|
| A1 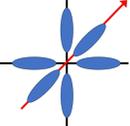 | Silicon full band $m_\sigma^*$ | Intra- **and** inter-valley | 0.51 | -2.3 |
| A2 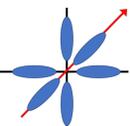 | Silicon full band $m_\sigma^*$ | Intra-valley **only** | 0.79 | -3.5 |
| B1 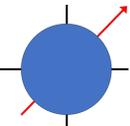 | Single spherical valley $m^* = m_\sigma^*$ | Intra- **and** inter-valley | 0.41 | -1.5 |
| B2 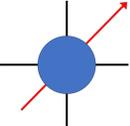 | Single spherical valley with $m^* = m_\sigma^*/6$ | Intra-valley **only** | 0.20 | 0 |

Table 1. Comparison of the b-factors as computed from a numerical solution with inter-valley scattering (A1) and without (A2) to corresponding spherical bands with (B1) and without (B2) inter-valley scattering. The extracted effective masses from the silicon band structure are $m_t = 0.22$ and $m_l = 0.93$ for transverse and longitudinal directions respectively, which gives $m_\sigma^* = 2.24 m_o$.

## 5. Discussion

Several authors have developed measures of Fermi surface complexity. For example, Mecholsky, et al. developed measures for the warped valences bands of Si [39], and Toberer discusses a shape factor [27]. Recently, Gibbs, et al. introduced a simple, numerical metric they call a Fermi Surface Complexity Factor (FSCF) [40]

$$FSCF = \left(\frac{m_s^*}{m_c^*}\right)^{3/2} = N_V^* K^*, \tag{17}$$

where $m_S^*$ is the so-called Seebeck effective mass (which is the density-of-states effective mass determined from the Seebeck coefficient) and $m_c^*$ is the conventional conductivity effective mass [35,41]. Metrics like this are useful in identifying promising thermoelectric materials in high-throughput searches of material databases [10, 42-47]. In (17), $K^*$ is an anisotropy factor that provides a measure of the benefits of Fermi surface anisotropy.

Equation (17) is an attempt to provide a simple numerical measure of how much better a complex band structure is as compared to a corresponding spherical, parabolic band. For a single valley with an effective mass of $m^*$, (17) gives $FSCF = 1$. For $N_V$ spherical valleys with an effective mass of $m^*$, (17) gives $FSCF = N_V$ (because $m_S^* \propto N_V^{2/3}$). Inserting numbers relevant for the conduction band of Si $\left(N_V = 6, m_t = 0.19 m_0, m_\ell = 0.93 m_0\right)$, we find $FSCF \approx 8.35$, which reflects both the benefits of a valley degeneracy of 6 and the valley anisotropy, which produces a light effective mass in the direction of transport (in this case, $K^* = 8.35/6 = 1.39$). Comparing case B2 to case A1 in Table 1, we see that in the presence of strong inter-valley scattering, $FSCF$ is an overly optimistic measure of the benefits of the anisotropic, multi-valley conduction band of Si. The anisotropy factor itself is a better indicator in this case, because the benefits of the additional valleys are offset by additional intervalley scattering.

The benefits of anisotropy in the presence of strong inter-valley scattering can be assessed by assuming that electron-phonon scattering follows the total DOS as given by (10). In this case, it is easy to show from (8f) to (8g) that the transport distribution simplifies to

$$\Xi(E) = \frac{\langle v_x^2(E) \rangle}{K_{el-ph}}. \tag{18}$$

Recall that the brackets indicate an average over angles at the energy, $E$. The fact that $v_x^2(E)$ plays an important role in thermoelectric performance has been noted before (e.g. in [5]). By energy-averaging the angle-averaged velocity squared, $\langle v_x^2(E) \rangle$, over the Fermi window, a metric sensitive to anisotropy and not to the number of valleys would result. Thus, $\langle v_x^2(E) \rangle$ could be a good measure of potential anisotropic enhancement.

## 6. Summary

Assessing the performance potential of a complex thermoelectric material involves a careful consideration of the number of valleys and bands that participate in transport, the role of scattering between these valleys and bands, and the effects of anisotropy. We illustrated how these issues can be examined by using n-type Si dominated by electron-phonon scattering as a model material. The calculations presented illustrate how rigorous treatments of electron scattering can inform calculations done by solving the Boltzmann Transport Equation in the Relaxation Time Approximation (RTA) as given by eqns. (8). For complex materials, these calculations can be computationally demanding, but they can address important questions such as the validity of the RTA, the use of energy-dependent rather than k-dependent scattering times, the relative strengths of intra- vs. inter-valley/band scattering, the energy dependence of the scattering time, etc.

The calculations presented here show that the degree to which multiple anisotropic valleys improve $zT$ depends sensitively on the relative strengths of intra- vs. inter-valley electron scattering processes. This fact is well known; the contribution of this paper is to show how this question can be quantitatively addressed. Anisotropy also plays an important role, and by assuming equally strong scattering within and between valleys, its effect can be assessed by a metric based on $\langle v_x^2(E) \rangle$. Widely used approximations, such as the constant relaxation time approximation (CRTA) and the constant mean-free-path approximation are not suitable for understanding the performance potential of a complex thermoelectric material with multiple valleys and bands. As illustrated in this paper, a combination of rigorous scattering calculations and standard RTA based solutions of the BTE may provide a more realistic way to assess the potential of complex materials.

**Supplementary Material**

See supplementary material for derivations of the anisotropic velocity ratio, distribution of Modes, density of states for spherical and ellipsoidal energy surfaces, and an alternate Fermi Surface Complexity Factor (FSCF) for multiple ellipsoidal bands.

**Acknowledgements**

This work was partially supported by the Defense Advanced Research Projects Agency (Award No. HR0011-15-2-0037). J. Maassen would like to acknowledge support from NSERC (Discovery Grant RGPIN-2016-04881).

**Appendix A**

This appendix presents a short derivation of the Landauer form of the transport distribution, (8g). Just one of the diagonal components will be derived here. Beginning with (8f) and assuming an energy-dependent scattering time, we find

$$\Xi_{xx}(E) \equiv \sum_{\vec{k}} v_x^2 \tau(E) \delta(E - E_k), \tag{A1}$$

which can be written as

$$\Xi_{xx}(E) \equiv \frac{\sum_{\vec{k}} v_x^2 \tau(E) \delta(E-E_k)}{\sum_{\vec{k}} |v_x| \delta(E-E_k)} \times \frac{\sum_{\vec{k}} |v_x| \delta(E-E_k)}{\sum_{\vec{k}} \delta(E-E_k)} \times \frac{\sum_{\vec{k}} \delta(E-E_k)}{\Omega}, \quad \text{(A2)}$$

where $\Omega$ is a normalization volume. The third factor in (A2) is recognized as the density of states,

$$D(E) \equiv \frac{1}{\Omega} \sum_{\vec{k}} \delta(E-E_k). \quad \text{(A3)}$$

The second factor in (A2) can be recognized as the angle-averaged velocity in the direction of transport,

$$\langle |v_x| \rangle \equiv \frac{\sum_{\vec{k}} |v_x| \delta(E-E_k)}{\sum_{\vec{k}} \delta(E-E_k)} \quad \text{(A4)}$$

The first term can be written as

$$\frac{\sum_{\vec{k}} v_x^2 \tau(E) \delta(E-E_k)}{\sum_{\vec{k}} |v_x| \delta(E-E_k)} = \frac{\sum_{\vec{k}} v_x^2 \tau(E) \delta(E-E_k)}{\sum_{\vec{k}} \delta(E-E_k)} \times \frac{\sum_{\vec{k}} \delta(E-E_k)}{\sum_{\vec{k}} |v_x| \delta(E-E_k)}$$

$$= \frac{\langle v_x^2 \tau_m \rangle}{\langle |v_x| \rangle} = \frac{\lambda(E)}{2} \quad \text{(A5)}$$

(see (9) in the text). Using (A3) – (A5) in (A2), we find

$$\Xi_{xx}(E) = \lambda(E) \frac{\langle |v_x| \rangle D(E)}{2}. \quad \text{(A6)}$$

Finally, using the definition for the number of channels per cross-sectional area [31, 34],

$$M(E)/A \equiv \frac{h}{4} \langle |v_x| \rangle D(E),$$

(A6) becomes

$$\Xi_{xx}(E) = \frac{2}{h} \lambda(E) M(E)/A. \quad \text{(A7)}$$

Equation (A7) expresses the transport distribution in terms of two physically clear factors, the mean-free-path for backscattering and the number of channels per cross-sectional area. The concept of channels is a seminal one in nanoscale transport, where $M(E)$ is a small countable number and leads to quantized conduction [48]. We use it here at a larger scale where $M(E)/A$ is large. Note also that the transport function is closely related to the transmission in the Landauer approach to transport [48].